\documentclass[%
 reprint,
showpacs,
 amsmath,amssymb,
 aps,
]{revtex4-1}

\usepackage{graphicx}
\usepackage{dcolumn}
\usepackage{bm}

\usepackage{xcolor}

\newcommand{\red}{\textcolor{black}}    

\usepackage[normalem]{ulem}

\begin{document}

\preprint{APS/123-QED}

\title{Electron localization and optical absorption of polygonal quantum rings}

\author{Anna Sitek}
 \email{sitek@hi.is}
 \affiliation{Science Institute, University of Iceland, Dunhaga 3, 
              IS-107 Reykjavik, Iceland}
 \affiliation{Department of Theoretical Physics, Wroc{\l}aw University of Technology, 
              50-370 Wroc{\l}aw, Poland}

\author{Lloren\c c Serra}
 \affiliation{Institute of Interdisciplinary Physics and Complex Systems  IFISC (CSIC-UIB), 
              E-07122 Palma de Mallorca, Spain}
 \affiliation{Department of Physics, University of the Balearic Islands, 
              E-07122 Palma de Mallorca, Spain} 
 
\author{Vidar Gudmundsson}
 \affiliation{Science Institute, University of Iceland, Dunhaga 3, 
              IS-107 Reykjavik, Iceland}

\author{Andrei Manolescu}
 \affiliation{School of Science and Engineering, Reykjavik University, 
              Menntavegur 1, IS-101 Reykjavik, Iceland}


\begin{abstract} 
We investigate theoretically
\red{polygonal quantum rings and focus mostly on the triangular geometry where the 
corner effects are maximal.  Such rings can be seen as short core-shell nanowires, 
a generation of semiconductor heterostructures with multiple applications. 
We show how the geometry of the sample determines the 
electronic energy spectrum, and also the localization of electrons, 
with effects on the optical absorption.}
In particular, we  show that irrespective of the ring shape 
low-energy electrons are always attracted by corners and are localized in 
their vicinity. The absorption spectrum in the presence of a magnetic field 
shows only two peaks within the corner-localized state domain, each associated 
with different circular polarization. This picture may be changed by an external 
electric field which allows previously forbidden transitions, and thus enables 
the number of corners to be determined. We show that polygonal quantum rings allow 
absorption of waves from distant ranges of the electromagnetic spectrum within one 
sample.  
\end{abstract}


\maketitle

\section{\label{sec:introduction} Introduction}

Recently it has become feasible to grow  core-multiple-shell nanowires 
consisting of a core built of one type of material which is surrounded
by one or more shells of different materials. This preparation method
makes achievable a huge variety of heterostructures with various
and controllable properties which make them extremely attractive as
building blocks of nanoelectronic and optoelectronic nanodevices,
in particular solar cells \cite{Krogstrup13,Pekoz11,Tang11,Wang14}
or nanoantennas \cite{Kim15}. 
\red{In particular, nanowires of triangular cross section turned out to
be a very good host for robust and efficient coaxial \textit{p-i-n} junctions
\cite{Dong09} or multicolor nanophotonic sources with controllable
wavelengths \cite{Gradecak05,Qian04,Qian05,Qian08}.
Besides these applications, we would also like to mention
a basic theoretical interest in polygonal rings as particular
examples of quantum graphs \cite{Berko} with characteristic physical
behaviors.}
Such nanowires are usually grown
vertically and, due to the crystallographic structure, have polygonal
cross sections, most commonly hexagonal \cite{Blomers13,Rieger12,Haas13},
\red{ but triangular \cite{Qian04,Qian05,Baird09,Dong09}, square
\cite{Fan06,Shtrikman09}}, and dodecagonal \cite{Rieger15} cross sections 
are also feasible. Sharp
edges along the wires induce unique carrier localization, which leads
to formation of one-dimensional (1D) channels in corner or side areas
\cite{Jadczak14, Bertoni11,Royo13,Royo14,Royo15,Fickenscher13, Shi15}.


Core-shell structures allow for modeling of many properties including
band alignment which strongly depends on the strain in the system and 
may be controlled through the core and/or shell thickness \cite{Pistol08}. 
In such a way one may grow systems in which electrons are confined only 
in the shell area \cite{Blomers13}. It is also possible to etch the core 
part and achieve hollow nanowires \cite{Rieger12, Haas13}, i.e., 
nanotubes of finite thickness. Multishell structures allow growth of 
narrow (up to $1.5$ nm) tubes which are formed between two shell layers 
such that surface effects are reduced \cite{Jadczak14, Fickenscher13, Shi15}. 
A polygonal nanoring may be considered as a short wire of this kind.  

Some insight has already been gained for hexagonal quantum rings which 
due to their symmetry and the possibility to localize electrons in the 
corners are refereed to as {\textit{artificial benzene}} \cite{Ballester12}.  
Electron localization at the corners of a polygonal quantum ring is expected 
if one notes that localization occurs whenever a nanowire is bent. Indeed, 
electronic states on nanowire bends, which in our case are the corners of 
the polygonal contour, attracted much interest some years ago 
\cite{Lent,Sols,Sprung,Wu,Wu2,Wu3,Vacek,Xu}. In a single-mode wire with a 
circular bend a simplified 1D picture was obtained in which the corner may 
be replaced by a square well, whose depth and length are determined by the 
angle and radius of the circular bend \cite{Sprung}.  This approach was 
used in Ref.\ \cite{Est} to suggest a scattering model of 1D polygonal wires, 
treating each corner as a scatterer. However, in contrast to our present 
purpose, the authors of Ref.\ \cite{Est} considered only the extended states along the 
sides of the 1D polygon. In fact, contrary to hexagonal quantum dot molecules, 
rings also allow localization of charge carriers in side areas, and they can 
even favor one of the sides if that is sufficiently thick \cite{Ballester12}. 
Other related effects, like the suppression of the Aharonov-Bohm effect in 
hexagonal rings, have also been theoretically envisioned \cite{Ballester13}.

In this paper we study electron localization in polygonal quantum rings of 
various shapes and show how it determines optical absorption.  We use a 
computational method based on finite differences on a polar grid which 
enables us to model not only hexagonal structures but arbitrary polygonal 
rings including nonsymmetric samples. We also derive the localized states 
with a 1D scattering model. 
\red{We focus mostly on triangular rings, where the corner-to-edge ratio is 
largest, resulting in the most pronounced localization and
corresponding optical effects.}
We show that irrespective of sample shape one can always distinguish a group of 
corner-localized states which for some geometries are separated by an energy 
gap from the higher states. The electron localization pattern is very sensitive
to sample symmetry and shape. For quantum rings defined by regular polygonal 
constraints the localization probability is equally distributed between all 
corners and/or sides of the ring, but when the symmetry is broken, e.g., by 
different side thicknesses or corner softening, the probability density becomes 
localized on individual corners which are occupied according to their areas. 
External electric fields may partially control electron localization, by 
destroying equal distribution between corners, or by delocalizing states 
previously occupying a single corner area.

We analyze optical absorption of the systems and show that in the presence 
of a perpendicular magnetic field only two transitions occur from the 
ground-state to corner- and side-localized domains, each associated with 
different polarization. Still, external electric fields may break the 
wave function symmetry such that more transitions become visible, and thus 
optical experiments may allow to infer the number of corners. Moreover, we 
point out that triangular quantum rings allow absorption in the microwave and 
near-infrared regimes to be observed within the same sample.

The paper is organized as follows. In Sec. \ref{sec:1dmod} we define the 
analyzed systems and make a preliminary inspection of the low-energy states 
in polygonal rings based on the 1D scattering model. Then, in 
Sec. \ref{sec:model} we introduce the sample model and describe the 
discretization method. In Sec. \ref{sec:estates} we present the low-energy 
quantum states resulting from our Hamiltonian model.  Then in 
Sec. \ref{sec:absorption} we calculate optical spectra corresponding to 
excitation of electrons initially in the ground-state.  Finally, 
Sec. \ref{sec:conclusions} contains conclusions and final remarks.

\section{\label{sec:1dmod} The 1D scattering model}

The systems under study are two-dimensional (2D) polygonal quantum rings of 
different shapes. They may also be considered as short core-multiple-shell or 
hollow nanowires such that all electronic wave functions include only the 
lowest axial mode. The first part of our analysis is based on a 1D scattering 
model.
A circular bend in a nanowire with a single transverse mode acts
approximately like a 1D square-well potential of depth $V_0$ and length
$2a$ \cite{Sprung}.  The bend radius $R$ and angle $2\theta$ determine
the effective square-well potential through the expressions $V_0\simeq
-\hbar^2/(8mR^2)$ and $a\simeq R\theta$. Such a potential always supports
bound states that, physically, represent states localized on the wire
bend \cite{Lent,Sols,Sprung,Wu,Wu2,Wu3,Vacek,Xu}.  This type of effective
confinement is the physical mechanism behind the  corner localization that
occurs in polygonal nanorings of finite width, which will be discussed
in detail in Sec. \ref{sec:estates}.

It is possible to devise a 1D  model, whose only coordinate is 
the position along the nanoring perimeter, say $\xi$, with cyclic 
boundary conditions on the wave function $\psi(\xi)$; namely $\psi(0)=\psi(L)$, 
where $L$ is the full perimeter length.
Each corner acts like a point scatterer, characterized by a scattering 
matrix given by a square well.
Using this 1D model the authors of Ref.\ \cite{Est} described the states propagating along 
the polygon sides. Here we extend that analysis to negative-energy states,
localized on the corners and behaving as evanescent waves on the polygon 
sides. 

The wave function between vertices $i$ and $i+1$ is a combination of 
right ($r$) and left ($l$) propagating plane waves,
\begin{eqnarray}
\psi(\xi) 
&=&
a_r^{(i)} e^{-ip(\xi-\xi_i)}
+
b_r^{(i)} e^{ip(\xi-\xi_i)}\nonumber\\
&=&
a_l^{(i+1)} e^{ip(\xi-\xi_{i+1})}
+
b_l^{(i+1)} e^{-ip(\xi-\xi_{i+1})}\; ,
\label{eq5}
\end{eqnarray}
where $i=1,\dots,N_v$ label the vertices, $\xi_i$ are their
positions, and the wave number $p$ is purely imaginary.  For a total
energy $E$ lower than the first transverse mode of the wire $\epsilon_1$
it is $p=i\sqrt{2m(\epsilon_1-E)}/\hbar$.  The scattering amplitudes
in Eq.\ (\ref{eq5}) fulfill a linear homogeneous system given by the
scattering relation
\begin{equation}
\label{ivs}
\left(
\begin{array}{c}
 b_l^{(i)}\\
\rule{0cm}{0.5cm} b_r^{(i)}
\end{array}
\right)
=
\left(
\begin{array}{cc}
  r & t \\
 t & r 
\end{array}
\right)
\left(
\begin{array}{c}
  a_l^{(i)}\\
\rule{0cm}{0.5cm}  a_r^{(i)}
\end{array}
\right)\; ,
\end{equation}
and the condition between successive vertices
\begin{eqnarray}
\label{bli}
b_l^{(i)} &=& a_r^{(i-1)} e^{-ip\ell}\; ,\\
\label{bri}
b_r^{(i)} &=& a_l^{(i+1)} e^{-ip\ell}\; .
\end{eqnarray}
In Eq.\ (\ref{ivs}) $r$ and $t$ are the reflection and transmission
scattering amplitudes of the above mentioned squared well for imaginary
wave numbers.  The energies $E$ for which Eqs.\ (\ref{ivs}), (\ref{bli}),
and (\ref{bri}) admit a solution can be determined from the zeros of
the determinant of the linear system matrix $M$, or equivalently, of 
the function \cite{Est}
\begin{equation}
\label{ff}
{\cal F}={\rm norm} \left\{ \tilde{M}
\left(
\begin{array}{c}
 a\\
b
\end{array}
\right)
\right\} \ ,
\end{equation}
where $\tilde{M}$ is analogous to $M$ except for an arbitrarily
chosen scattering amplitude which is set to 1.

\begin{figure}
\includegraphics[width=0.48\textwidth,angle=0]{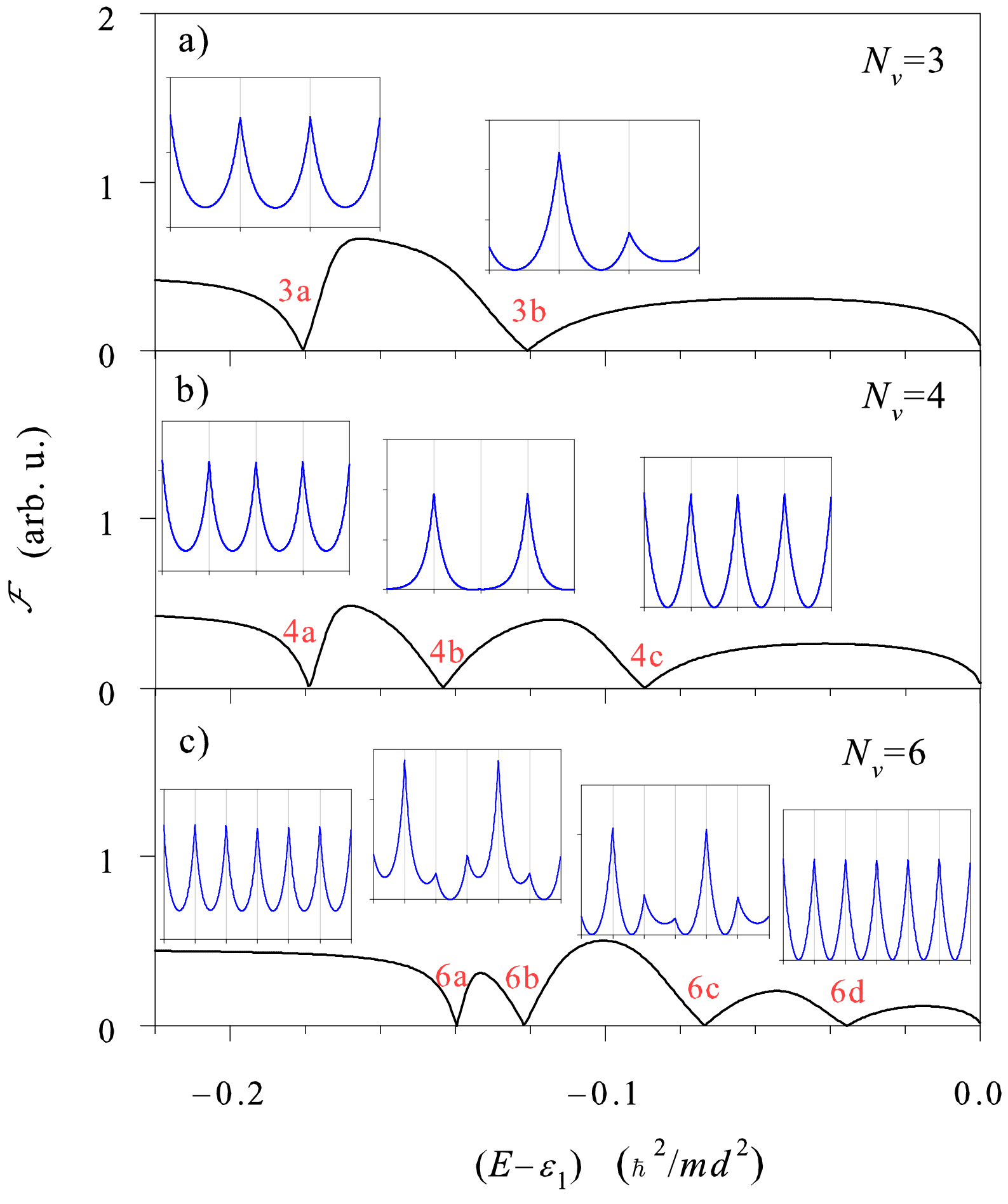}
\caption{${\cal F}$ function of the 1D model whose zeros signal the allowed states.
The insets show the corresponding 1D densities for each  zero ${\cal F}$
with the vertical lines corresponding to the vertex positions.
The number of vertices $N_v$ is indicated in each panel.
In terms of the width $d$, the side length is fixed at  
$\ell=5d$ and the bend radii are $R=0.15d$, $0.1d$, and $0.07d$ for the 
triangle, square, and hexagon, respectively.}
\label{1dfig}
\end{figure}

Figure \ref{1dfig} shows the energy dependence of ${\cal F}$ for a
triangular, a square, and a hexagonal sample. The sequence of allowed 
energies in
each polygonal nanoring is seen from the ${\cal F}$ zeros while the
figure insets show the corresponding 1D densities for each mode (labeled
as $N_v$a, $N_v$b, etc.).  
The probability densities
are concentrated on the corners and the modes can be classified into
two types: translational symmetric modes (TSM's) having the
same density on each segment of the polygon and translational
asymmetric modes (TAM's) for which the sides look different. 
The TAM's (3b, 4b, 6b, 6c)
are degenerate, since inversion from the central $\xi$ point leads to
another valid solution.  Counting also the spin, the degeneracy factors
become 2 for symmetric modes and 4 for asymmetric ones.

A closer look at the TSM's of Fig.\ \ref{1dfig} reveals that
there are two types, depending on the density at each side
midpoint. Modes 3a, 4a, and 6a have finite midpoint densities while modes
4c and 6d exactly vanish at midpoints.  With a similar analysis as that
of Ref.\ \cite{Est} it can be shown that the first type (3a, 4a, 6a)
occurs when $t+r=e^{-ipl}$. The second type of TSM's
(4c, 6d) correspond to the condition $t+r=-e^{-ipl}$ and occur only in
even $N_v$ polygons.  We also notice that the energies of the
TAM's lie in between the TSM's. The sequence is such that
in odd-$N_v$ polygons there are $N_v-1$ TAM's between TSM's, 
while in even-$N_v$ polygons there are $N_v/2-1$ intermediate
TAM's.  In all cases, however, this sequence of localized states abruptly
terminates at the side-propagating threshold $E=\epsilon_1$.

\section{\label{sec:model} The Hamiltonian model}

\begin{figure}
\includegraphics[width=0.21\textwidth,angle=0]{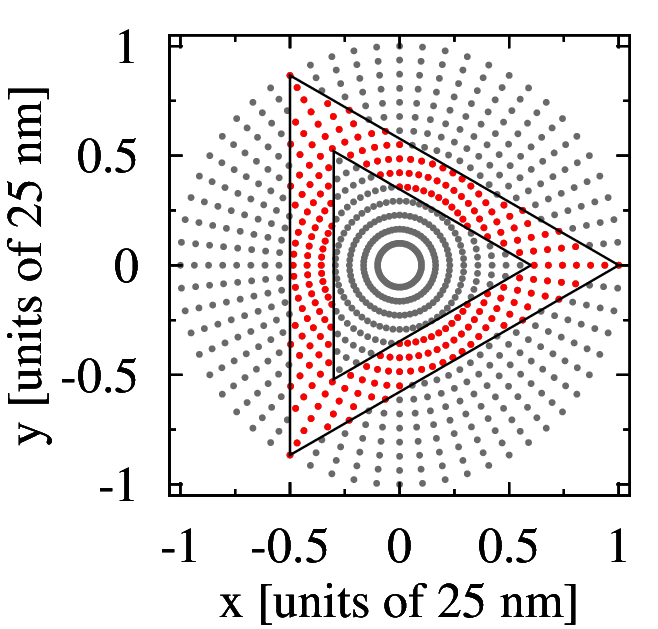}
 \caption{Sample model: Polygonal constraints applied on a polar grid. 
          For visibility we reduced the number of site points.}
 \label{Fig_Sample}
\end{figure}


The second part of our
modeling is based on a discretization method on a polar grid. We
start with a circular disk geometry as in Ref.\ \onlinecite{Daday11}
on which we apply polygonal constraints and pick up only points within
the resulting shell (Fig.\ \ref{Fig_Sample}). In this case the Hilbert
space is spanned by vectors $|kj\sigma\rangle$, where $k$ and $j$ label
the discretized radial and angular coordinates $(r_k, \phi_j)$,
with meshes ($\delta r, \delta \phi$), respectively, and $\sigma$
stands for the two possible spin values.

The system Hamiltonian consists of four terms,
\begin{eqnarray}
\label{Hamiltonian_1}
H = H^K+H^E+H^B+H^Z.
\end{eqnarray}
Hamiltonian matrix elements of the first contribution, the kinetic Hamiltonian, 
in the $k$, $j$, and $\sigma$ 
basis are
\begin{eqnarray}
\label{Hamiltonian}
H^{K}_{kj\sigma,k'j'\sigma'} = 
T\delta_{\sigma,\sigma'}\left[ t_r \left(\delta_{k,k'}
-\delta_{k,k'+1}\right)\delta_{j,j'}\right. \nonumber \\ 
\left. + t_{\phi}\delta_{k,k'}\left(\delta_{j,j'}-\delta_{j,j'+1}\right) 
+ \mathrm{H.c.}\right], 
\end{eqnarray}
where $T=\hbar^{2}/(2m^{*}R^{2}_{\mathrm{ext}})$ is a reference energy, 
$m^{*}$ is the effective mass of the semiconductor material,
$R_{\mathrm{ext}}$ is the external radius of the polar grid, 
$t_r=(R_{\mathrm{ext}}/\delta r)^2$, and
$t_{\phi}=[R_{\mathrm{ext}}/(r_k\delta\phi)]^2$.

We expose the rings to external electric and magnetic fields. The electric
field is parallel to the $x$-$y$ plane 
and forms an angle $\varphi$ with the $x$ axis, 
$\bm{E}=E(\cos\varphi, \sin\varphi, 0)$, and 
the corresponding Hamiltonian matrix elements are 
\begin{eqnarray*}
\label{Hamiltonian_K}
H^E_{kj\sigma,k'j'\sigma'} = 
- e {\bm E} \cdot {\bm r}_{k}
\delta_{k,k'}\delta_{j,j'}\delta_{\sigma,\sigma'},
\end{eqnarray*}
where $e$ is the electron charge.
The magnetic field ${\bm B}$ is assumed perpendicular to the ring plane, 
with a vector potential $\bm{A}=B(-y,x,0)/2$, and the corresponding 
Hamiltonian matrix elements are obtained as
\begin{eqnarray*}
\label{Hamiltonian_B}
H^B_{kj\sigma,k'j'\sigma'} = 
T\delta_{\sigma,\sigma'} \delta_{k,k'} 
\left[\frac{1}{2}t^2_{\mathrm{B}}\left(\frac{r_k}{4R_{\mathrm{ext}}}\right)^2\delta_{j,j'} \right. \nonumber \\
\left. -t_{\mathrm{B}}\frac{i}{4\delta\phi}\delta_{j,j'+1} + \mathrm{H.c.}\right],
\end{eqnarray*}
with $t_{\mathrm{B}}=\hbar e B/ m^{*}T$ the cyclotron energy in units of $T$.

The last contribution to the Hamiltonian, the Zeeman part, is diagonal in 
the $k,j$ and $\sigma$ basis
\begin{eqnarray}
\label{Hamiltonian_Z}
H^{\mathrm{Z}}_{kj\sigma,k'j'\sigma'} = 
\frac{1}{2}Tt_{\mathrm{B}}\gamma\left(\sigma_z\right)_{\sigma,\sigma'}\delta_{k,k'}\delta_{j,j'},
\end{eqnarray}
where $\gamma=g^{*}m^{*}/2m_{e}$ is the ratio between the Zeeman gap 
and the cyclotron energy, and $m_e$ is the free-electron mass.
\red{Our discretization method is a version of the very popular
hopping schemes used in the mesoscopic physics. Other theoretical
studies of core-shell polygonal systems used the finite-element method
\cite{Fickenscher13, Wong11, Royo15}.}

\section{\label{sec:estates} Electronic states}

Below we present results for 2D polygonal rings achieved with the 
discretization method where the sample consists of over 6000 grid 
points. We use the external radius $R_{\mathrm{ext}}=25$ nm.  
We perform  numerical calculations for InAs parameters which are 
$m^{*}=0.023m_{e}$, where $m_{e}$ is the electron mass, and 
$g^{*}=-14.9$; thus the energy unit $T$ introduced in the 
Hamiltonian (\ref{Hamiltonian}) equals approximately $2.8$ meV and 
the ratio $\gamma=-0.171$.

\subsection{\label{sec:symmetric} Symmetric samples}

\begin{figure}
\includegraphics[width=0.48\textwidth,angle=0]{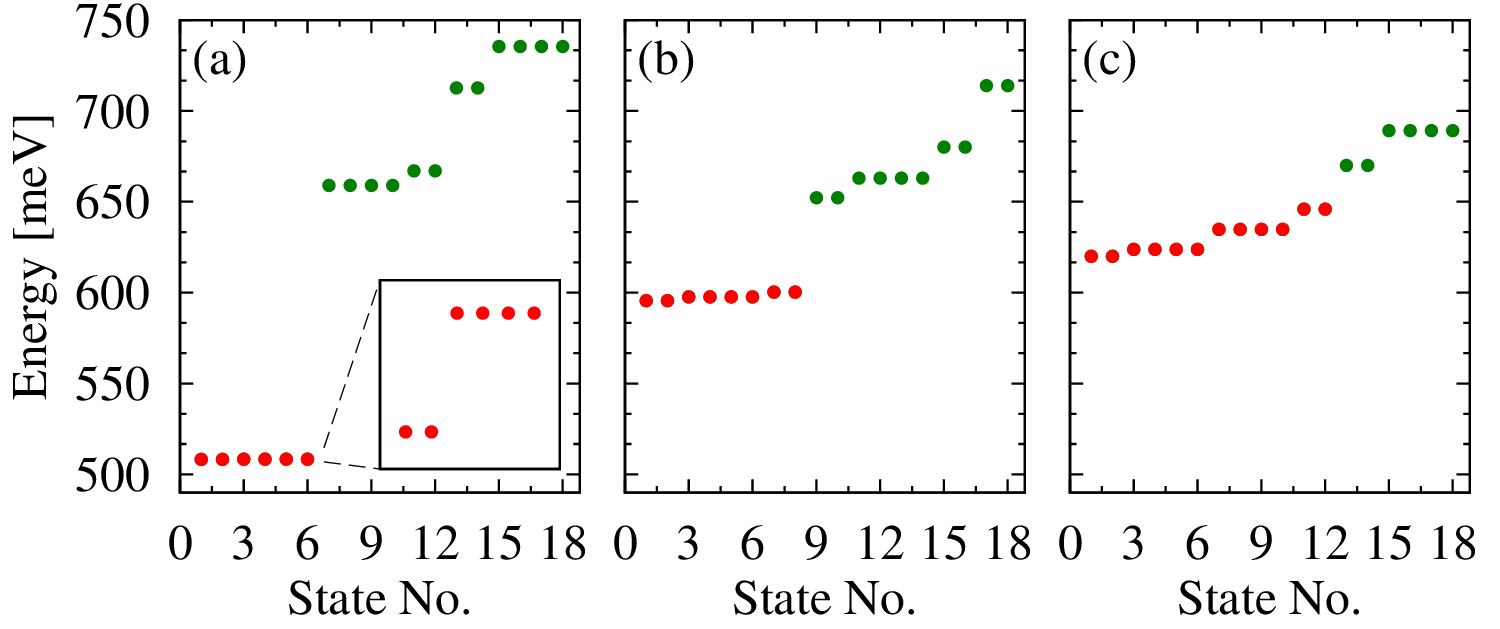}
 \caption{Energy levels for symmetric triangular (a), square (b), and 
          hexagonal (c) samples, of external radii equal to $25$ nm and side 
          thicknesses $5$ nm. The inset to (a) shows degeneracy of 
          the two lowest energy levels of the triangular sample. Red points 
          indicate purely corner-localized states.}
 \label{Fig_Energy}
\end{figure}


\begin{figure}
\includegraphics[width=0.48\textwidth,angle=0]{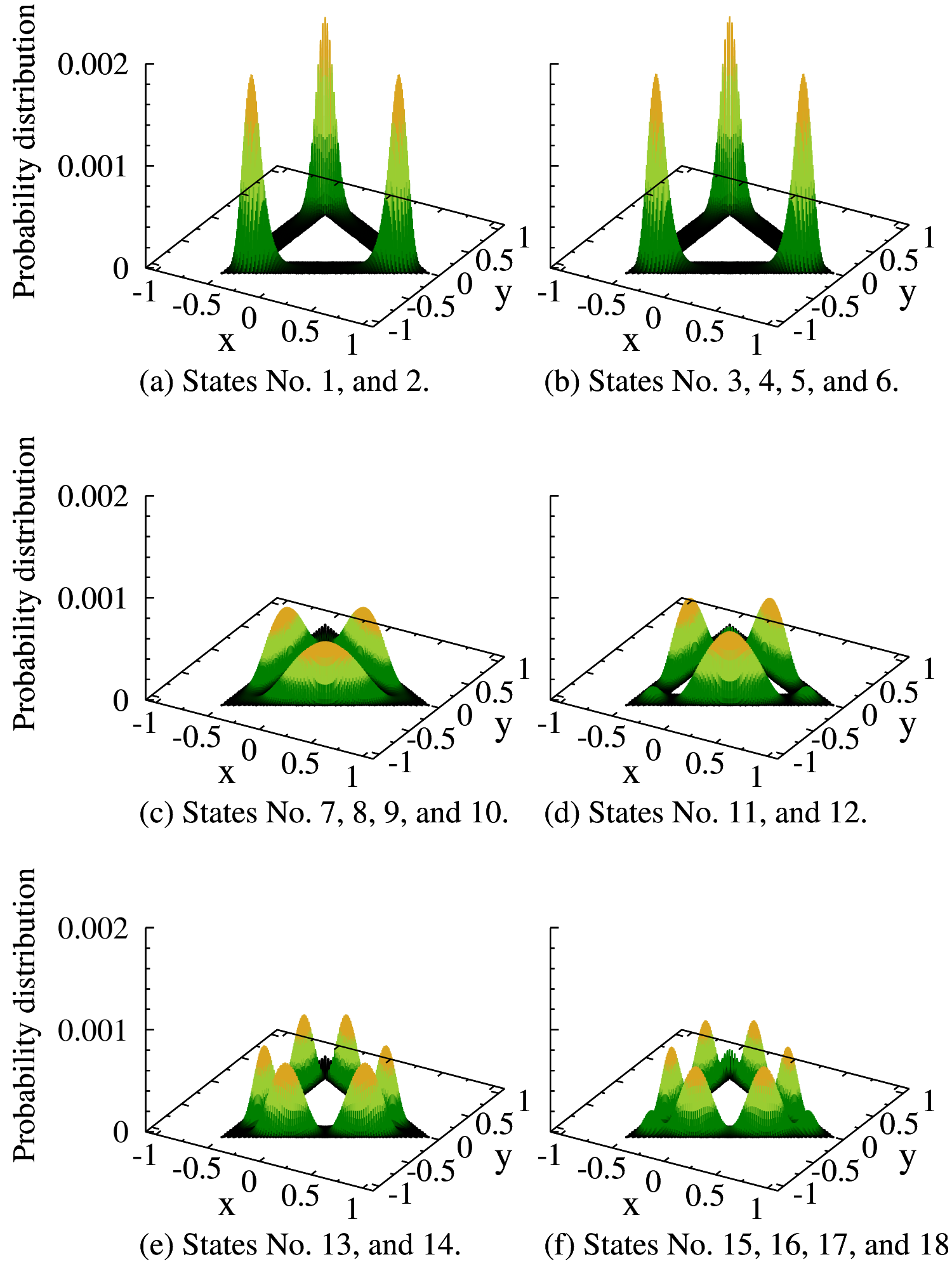}
 \caption{Probability distribution for the 18 lowest states of a symmetric 
          triangular ring. In (a) and (b) we show localization of the lowest 
          states indicated in red in Fig.\ \ref{Fig_Energy}(a), and in
          (c), (d), (e), and (f) localization of the states above 
          the energy gap in Fig.\ \ref{Fig_Energy}(a).}
 \label{Fig_prob_sym_tr}
\end{figure}


Symmetric polygonal samples which are restricted externally and internally
by regular polygons have well defined symmetries which imply specific energy 
degeneracies. In Fig.\ \ref{Fig_Energy} we compare the energy levels of a single 
electron confined in symmetric triangular, square, and hexagonal rings, all 
having sharp corners and $5$ nm side thicknesses, in the absence of external 
fields. As can be seen, the ground-state energy increases with the number of 
corners. This is because the size of the effective well formed in the corner 
area decreases with increasing corner angle, and thus ground-state electrons 
bounded in $2\pi/3$ corners of the regular hexagon have higher energy than those 
trapped in $\pi/3$ corners of regular triangles. This is in a nice qualitative 
agreement with the results shown in Fig.\ \ref{1dfig}.  
In a circular nanoring the ground-state has zero angular momentum and it is 
doubly (spin) degenerate, whereas all higher states are fourfold degenerate, 
having finite angular momenta that do not distinguish energetically between 
clockwise and counterclockwise electron rotations \cite{Fuhrer01, Aichinger06,
Nita11}.  When the regular $N_v$ polygonal constraints are applied to a ring
structure they break the circular degeneracy at levels corresponding to multiples 
of $2N_v$. The resulting series of two- and fourfold degenerate energy levels 
agree with the expectation from Sec. \ref{sec:1dmod} where spin was ignored.

In Fig.\ \ref{Fig_Energy}(a) a group of the six lowest states of the triangular
ring is separated from the higher states. The energy gap behind the 
eighth state is still visible for a square polygon [Fig.\ \ref{Fig_Energy}(b)], but 
considerably decreased with respect to the triangular sample, and it practically 
vanishes for a hexagonal ring [Fig. \ \ref{Fig_Energy}(c)]. Although the energy 
spacing between the $12$th and $13$th states of the \textit{artificial benzene} 
is comparable with other energy differences, in this case also the lowest states 
have different character from the others. States associated with the lowest
energy levels of symmetric polygonal rings (red points in Fig.\ \ref{Fig_Energy}) 
are equally distributed between all of the corners, as is shown for a triangular
ring in Figs.\ \ref{Fig_prob_sym_tr}(a), and \ \ref{Fig_prob_sym_tr}(b). Due to 
the spin degeneracy the number of these states equals double the number of 
corners ($2N_v$). If the sample is thick enough and contains a sufficient number 
of corners the probability distribution does not vanish completely in the middle 
of the sides, as for the triangular ring shown in Fig. \ref{Fig_prob_sym_tr}, but 
stabilizes at a much lower level than the corner maxima. The first state above the 
corner states is purely localized in sides, with maximal probability of finding a 
particle in the middle of each side [Fig.\ \ref{Fig_prob_sym_tr}(c)]; higher energy 
electrons are also mostly localized in the side areas with only a small probability of 
finding them in corners [Fig.\ \ref{Fig_prob_sym_tr}(d)]. The number of probability
maxima in the side regions increases with energy and the possibility of finding them 
in corners becomes relevant [Figs.\ \ref{Fig_prob_sym_tr}(e), 
and \ \ref{Fig_prob_sym_tr}(f)]. The probability of finding electrons in sharp 
corners becomes comparable to or even exceeds side maxima for high-energy electrons, 
but the detailed analysis of such states is beyond the scope of this paper.

For the lowest, corner-localized states, the probability density maxima decrease 
with increasing number of corners and at the same time the density of localization 
areas increases, similarly, for the first state above them the number of maxima 
increases and the distances between them decrease with an increasing number of 
corners, i.e., the side localization areas decrease. As a result the probability 
distributions for corner- and side-localized states become relatively similar and 
thus the energy gap occurring for triangular and square quantum rings 
[Figs.\ \ref{Fig_Energy}(a) and \ \ref{Fig_Energy}(b)] vanishes for sufficiently 
thick hexagonal samples [Fig. \ \ref{Fig_Energy}(c)]. However, the corner-localized 
states of {\textit{artificial benzene} may also be energetically separated from the higher 
states when the rings are very narrow such that the corner-localization areas are 
much smaller than the side ones.

\begin{figure}
\includegraphics[width=0.37\textwidth,angle=0]{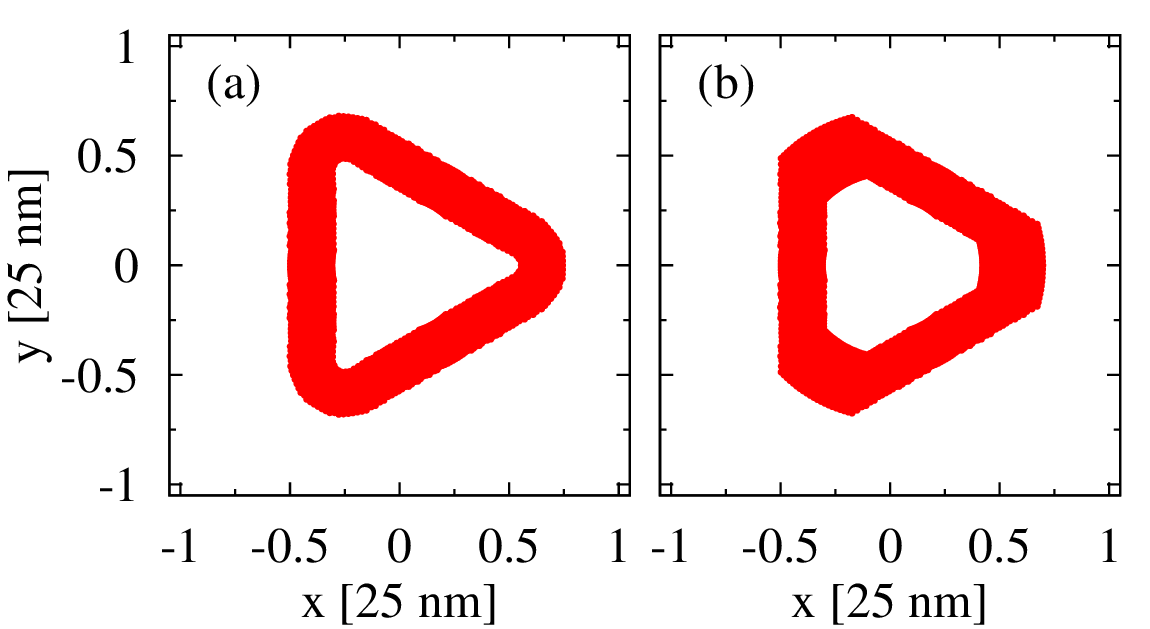}
 \caption{Triangular quantum rings with softened corners. (a) Round corners 
          softened by circles of 
          $r_{\mathrm{int}}=0.05R_{\mathrm{ext}}$ and $r_{\mathrm{ext}}=0.25R_{\mathrm{ext}}$ 
          inscribed in the internal and external limiting polygon corners. (b) Corners 
          softened by background ring radii reduced to $70\%$ of the 
          distance from the center of the sample to the sharp internal and 
          external corners.}
 \label{Fig_Sample_soft}
\end{figure}


\begin{figure}
\includegraphics[width=0.48\textwidth,angle=0]{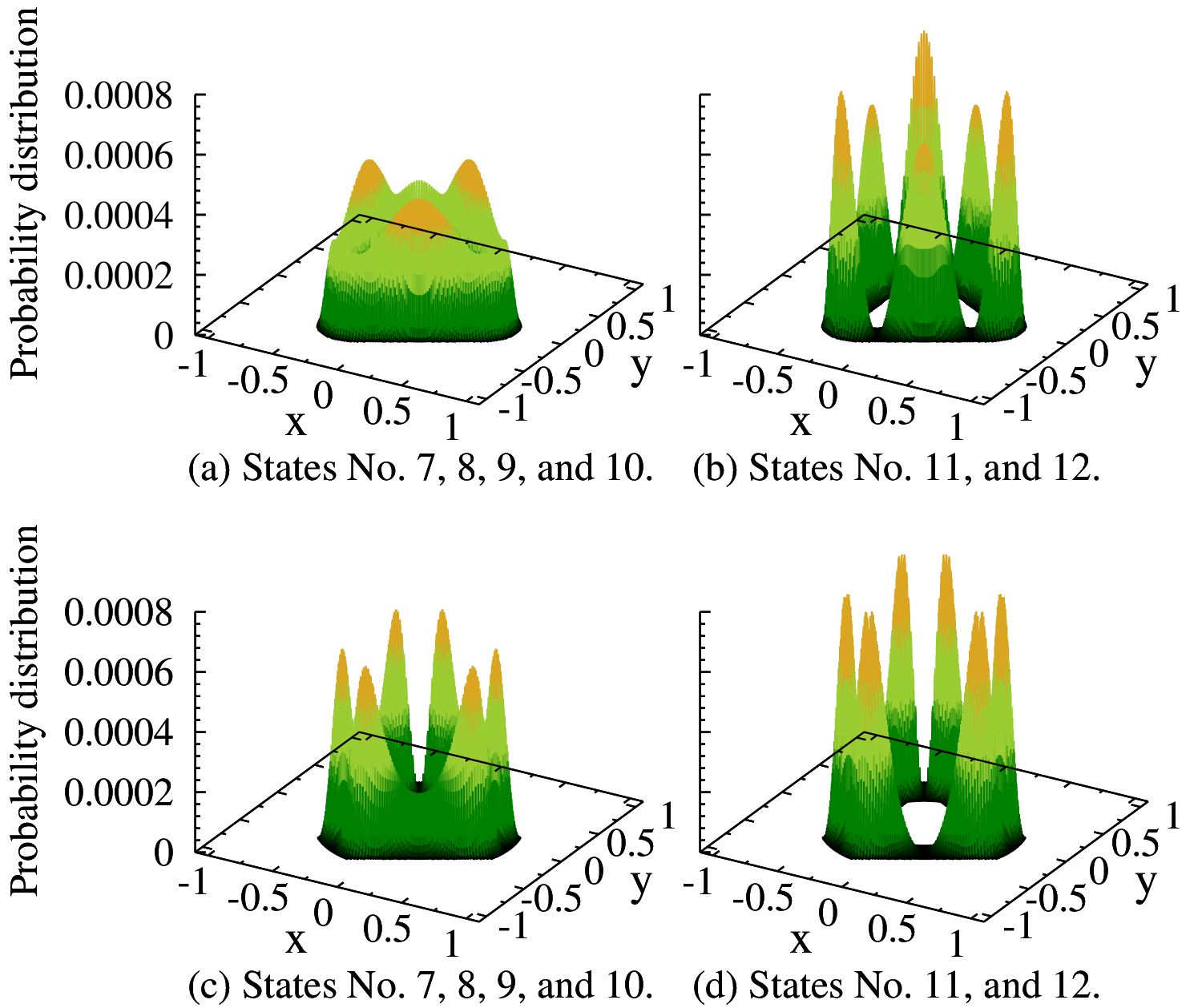}
 \caption{Probability distribution for the states associated with the 
          third (a) and (c) and fourth (b) and (d) energy levels of 
          the soft-corner samples. (a) and (b) refer to the sample 
          shown in Fig.\ \ref{Fig_Sample_soft}(a), and (c) and (d)
          to the sample in Fig.\ \ref{Fig_Sample_soft}(b).}
 \label{Fig_prob_soft}
\end{figure}


In practice it may be difficult to achieve samples with perfectly sharp 
corners. Therefore we investigated the impact of corner softening on energy 
levels and carrier localizations. We analyze two types of symmetric 
triangular samples shown in Fig.\ \ref{Fig_Sample_soft}, in one case we 
inscribe circles in the corners which define new internal and external 
limits in corner areas [Fig.\ \ref{Fig_Sample_soft}(a)]; in the other case 
we soften corners by ´cutting´ the sharp parts by background radii 
[Fig.\ \ref{Fig_Sample_soft}(b)]. In both cases energy levels show the 
same degeneracies as for the samples with sharp corners 
[Fig.\ \ref{Fig_Energy}(a)]. Moreover, the lowest six states associated with
the two lowest energy levels are always localized in corner areas. If all
of the corners are equally softened and when the softening is relatively
small, which for $5$-nm-thick samples means that the radii reduction for
the sample shown in Fig.\ \ref{Fig_Sample_soft}(b) must be up to around
$80\%$, then the probability density for samples with soft corners does
not differ considerably from the one shown in Fig.\ \ref{Fig_prob_sym_tr}. 
There are many possibilities of softening internal and external corners
separately; thus there is a huge variety of samples which show properties
of ideal (sharp) ones. Interesting features appear when these limits are 
exceeded. In the case of the sample shown in Fig.\ \ref{Fig_Sample_soft}(a) 
the energy gap separating the purely corner-localized states is comparable 
with energy splittings occurring \red{in the next higher states. 
Those states can be distributed both in the corners and on the sides
[Figs.\ \ref{Fig_prob_soft}(a) and \ \ref{Fig_prob_soft}(b)].}


The energy separating the two lowest energy levels exists in samples with
´cut´ corners, like the one shown in Fig.\ \ref{Fig_Sample_soft}(b),
but it decreases \red{due to} softening.  In contrast to the samples with 
sharp corners, where the lowest six states are localized in the corners 
and the next six ($7$th to $13$th) states are localized on the sides of the 
triangle (Fig. \ref{Fig_prob_sym_tr}), now the states associated with the 
two levels above the energy gap ($7$th to $13$th states) are still localized 
in the corner area, but each one has two nearby maxima. In fact corner 
softening of this kind increases \red{the number of corners, and the sample may 
show mixed features of triangles and hexagons.  Three corner maxima split in each 
corner area such that six maxima are formed [Figs.\ \ref{Fig_prob_soft}(c) 
and \ \ref{Fig_prob_soft}(d)] and the transition to mostly (not purely) 
side-localized states occurs above the fourth energy level as for hexagonal
samples.}


\subsection{\label{sec:non_symmetric} Nonsymmetric samples}

\begin{figure}
\includegraphics[width=0.262\textwidth,angle=0]{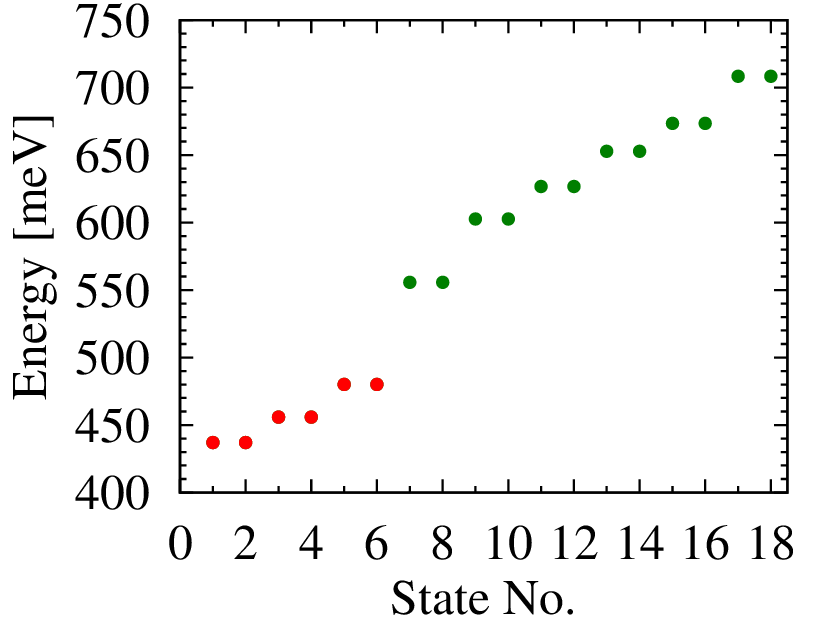}
 \caption{Energy levels for a nonsymmetric triangular ring where the 
          symmetry was broken by increasing the side thickness by $5\%$ 
          (side parallel to the $y$ axis) and $10\%$ (subsequent side according 
          to counterclockwise counting). Red points indicate corner-localized 
          states.}
 \label{Fig_Energy_non}
\end{figure}


\begin{figure}
\includegraphics[width=0.48\textwidth,angle=0]{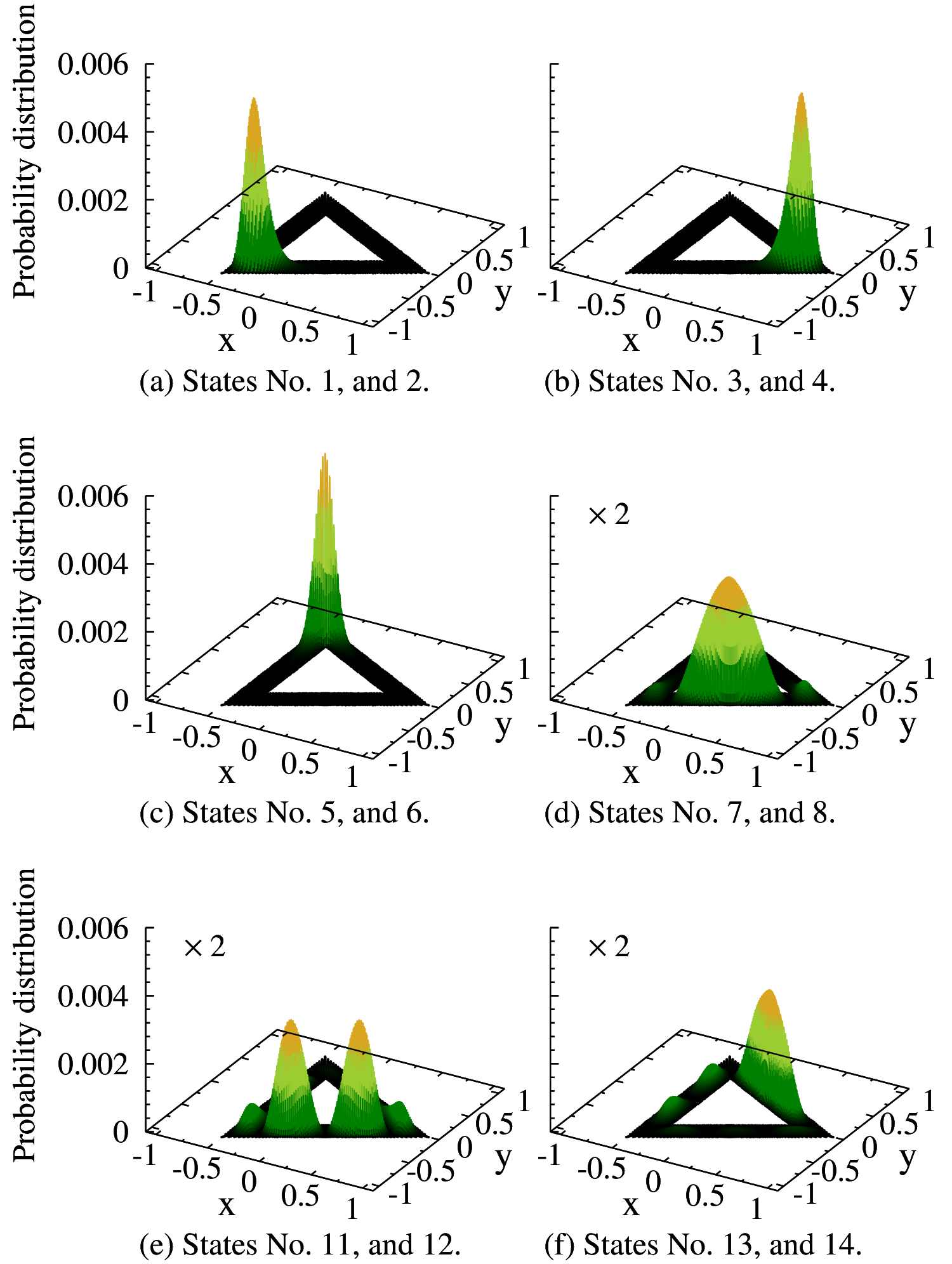}
 \caption{Probability distribution for a few chosen energy levels of 
          the nonsymmetric sample shown in Fig.\ \ref{Fig_Energy_non}. 
          Results shown in (d), (e), and (f) have been 
          graphically magnified by a factor of $2$ ($\times 2$).} 
          \label{Fig_prob_non_tr}
\end{figure}


Although the present state of the art of manufacturing allows high precision control
at the single-atom level, it is still difficult to grow perfectly
symmetric nanowires, and thus we also analyze different nonsymmetric
samples. First we break the symmetry by increasing the thickness of two
sides by $5\%$ and $10\%$. In this case the energy levels are only spin
degenerate and the energy gap between the sixth and seventh states is
reduced with respect to the symmetric case, but it is still relevant
(Fig.\ \ref{Fig_Energy_non}). The lowest states are also localized in
corner areas (red points in Fig.\ \ref{Fig_Energy_non}), 
but the probability distribution is not spread on all
corners as before: electrons of specific energy values
occupy only one corner. In particular, the ground-state is localized
in the corner with the largest area [Fig.\ \ref{Fig_prob_non_tr}(a)],
the two states associated with the second energy level occupy the corner
with the medium area [Fig.\ \ref{Fig_prob_non_tr}(b)], and the electrons
possessing the third energy value may be found in the smallest corner,
Fig.\ \ref{Fig_prob_non_tr}(c). The first state above the corner
localized group is mostly localized in the side region with small
probability peaks in corners [Fig.\ \ref{Fig_prob_non_tr}(d)]. In
general the number of probability peaks increases with the
energy. But unlike what happens in a symmetric polygon, one can obtain states
with probability distribution concentrated in fewer places than in
lower-energy states, as is shown in Figs.\ \ref{Fig_prob_non_tr}(e) and \
\ref{Fig_prob_non_tr}(f). Similar situations (not shown) occur for
rings which are defined by nonsymmetric polygons with uniform 
thicknesses, but  nonuniform angles. Since the wells formed in corner 
areas of polygonal rings depend on the angle geometry, they become 
nonsymmetric when the polygonal sides have different thicknesses. Thus 
when one side of the polygon is thicker than the others then the wells formed 
at its ends have the largest areas and are shifted towards the center of 
the thicker side. This results in delocalization for the ground-state 
electrons in those corner areas and shifting them towards the center of the 
widest side with increasing side thickness. For sufficiently thick 
sides the wells merge, and the ground-state becomes localized in the middle 
of the thicker side, as shown in Ref.\ \onlinecite{Ballester12}.

\begin{figure}
\includegraphics[width=0.48\textwidth,angle=0]{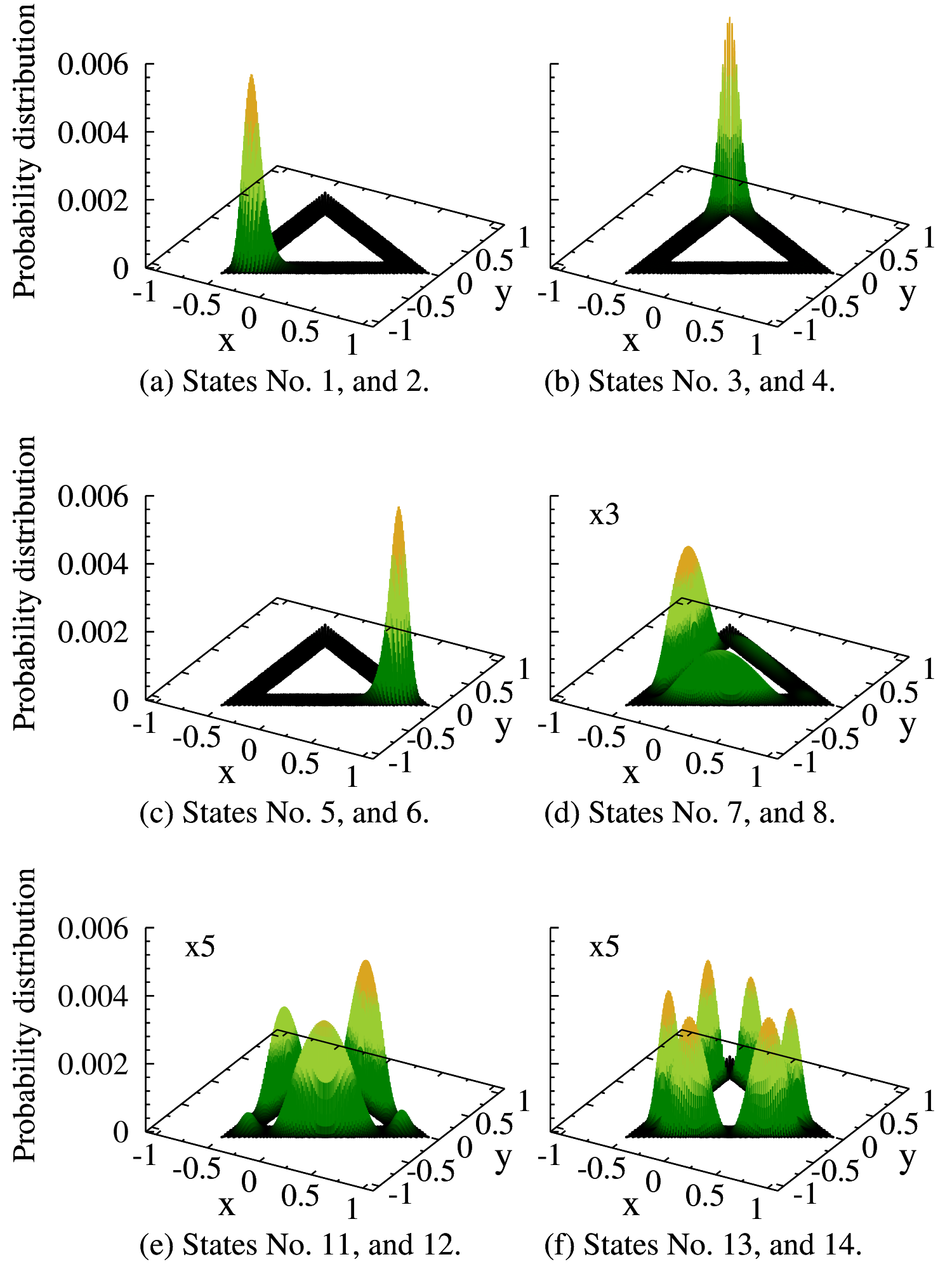}
 \caption{Probability distribution for a few chosen energy levels of 
          a symmetric sample in the presence of an electric field applied 
          along one of the sides \red{($E\approx 0.11$ mV/nm, $\varphi=\pi/6$)}.          
          Results shown in (d), (e), and (f) have been 
          graphically magnified by a factor of $3$ or $5$ 
          ($\times 3$, $\times 5$).} 
 \label{Fig_prob_el_sym}
\end{figure}


\begin{figure}
\includegraphics[width=0.48\textwidth,angle=0]{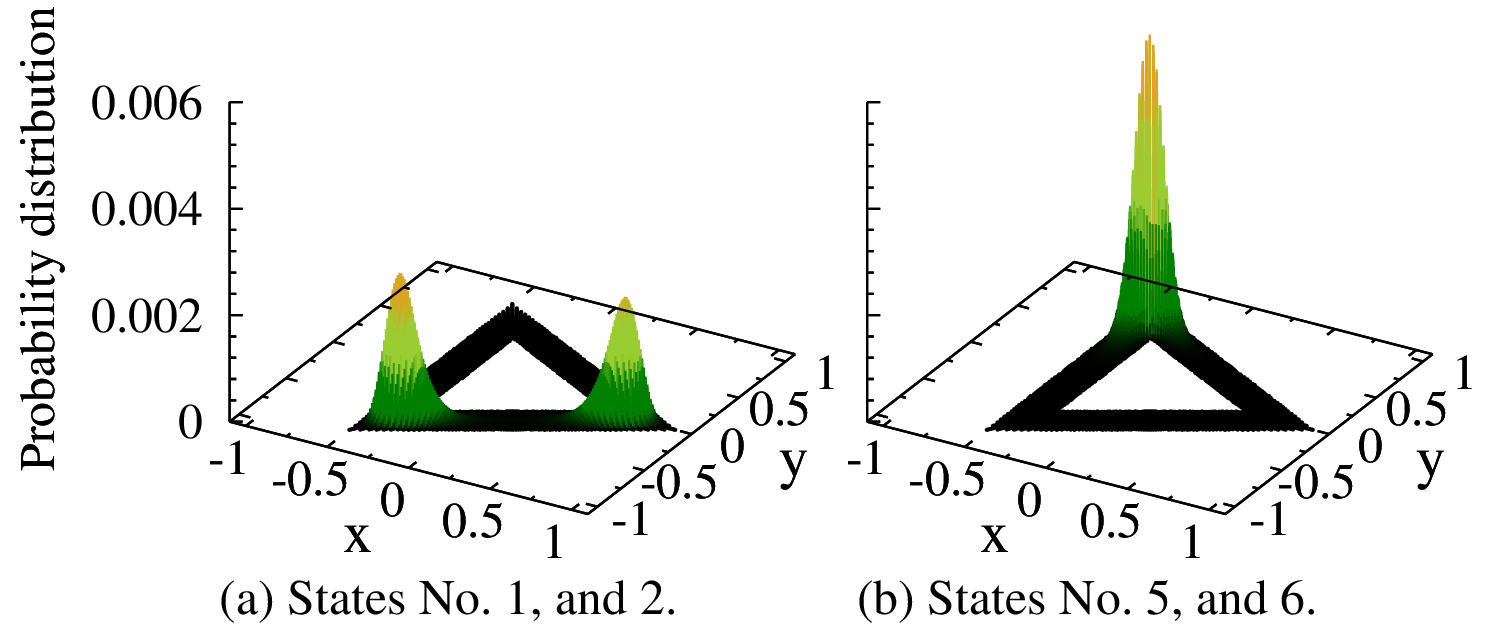}
 \caption{Probability distribution associated with the ground-state 
          (a) and with the third energy level (b) for a nonsymmetric 
          sample in the presence of an electric field 
          \red{($E\approx 0.67$ mV/nm, $\varphi=\pi/6$)}.
          }
 \label{Fig_prob_el_non}
\end{figure}


The electron probability density may be controlled externally by applying
electric fields. On one hand this may break the symmetric distribution
of regular samples; on the other hand it may rebuild, if not perfectly then to
some extent, a symmetric distribution in nonsymmetric samples. In Fig.\
\ref{Fig_prob_el_sym} we show electron localization for a geometrically
symmetric sample in the presence of an external electric field parallel
to one of the sides (forming an angle $\pi/6$ with the $x$ axis).
The electron localization resembles much more that of the nonsymmetric
triangle (Fig.\ \ref{Fig_prob_non_tr}) than that of the regular one (Fig.\
\ref{Fig_prob_sym_tr}). But, in contrast to the case with different side
thicknesses, here one can change the order of corner occupation. For
example, with the electric field rotated such that it becomes parallel
to the $y$ axis, the ground-state would be localized as in Fig.\
\ref{Fig_prob_el_sym}(a), but the corner areas associated with the
two higher energy levels would be reversed with respect to Fig.\
\ref{Fig_prob_el_sym}. Or the field perpendicular to one side may
localize the ground-state in the opposite corner, whereas the two higher
(nearly degenerate) energy levels become equally distributed between
the remaining two corners.

If the previously described sample with different side thicknesses
is placed in an external electric field, then it is quite easy to
delocalize the ground and the first excited states between the corners
at the ends of the widest side [Fig.\ \ref{Fig_prob_el_non}(a)] while 
the probability distribution of the highest corner-localized state remains
localized in the smallest corner [Fig.\ \ref{Fig_prob_el_non}(b)].
Restoration of equal distribution between all three corners is impossible
due to the particular shapes and areas of the quantum wells formed in the 
corners, which strongly differ from each other and cannot be compensated 
by an external electric field for any angle $\varphi$.


\section{\label{sec:absorption} Optical absorption}

We describe the interaction of electrons in the polygonal rings with
an external radiation field in the dipole approximation. The optical
absorption coefficient at zero temperature is given by the well-known
formula \cite{Haug09,Chuang95,Hu00}
\begin{eqnarray}
\label{absorption_coeff}
 \alpha(\hbar\omega) = A\hbar\omega\sum_{\mathrm{f}}|\langle f|\bm{\varepsilon}\cdot\bm{d}|i\rangle|^2
 \delta\left(\hbar\omega - \left(E_{\mathrm{f}}-E_{\mathrm{i}}\right)\right),
\end{eqnarray}
with $A$ being a constant containing physical parameters such as the 
refractive index, the speed of light, the dielectric permittivity, and the 
sample area, $\bm{\varepsilon}=\left(1,\pm i\right)/\sqrt{2}$ the circular 
photon polarization, $\bm{d}$ the dipole moment, and $E_{\mathrm{i,f}}$ the
energies of the initial and final states $|i,f\rangle$, respectively. 
The dipole matrix elements are
\begin{eqnarray*}
\langle f|\bm{\varepsilon}\cdot\bm{d}|i\rangle = \frac{1}{\sqrt{2}}\sum_{q}
\Psi^{\dagger}\left(q,f\right)\Psi\left(q,i\right)r_q\left(\cos\phi_q\pm i \sin\phi_q\right),
\end{eqnarray*}
where the summation is carried out 
over all possible basis states $|q\rangle\equiv |kj\sigma\rangle$ and $\Psi(q,a)$ 
are the amplitudes of the eigenvectors of Hamiltonian (\ref{Hamiltonian_1}) 
in the $q$ basis, $|a\rangle=\sum_q\Psi\left(q,a\right)|q\rangle$, 
where $H|a\rangle=E_a|a\rangle$. We 
approximate the $\delta$ function by the spectral weight in the presence of a 
constant self-energy $\Gamma/2$, 
\begin{eqnarray}
 \delta\left(\hbar\omega -
 \left(E_{\mathrm{f}}-E_{\mathrm{i}}\right)\right)\approx
 \frac{\Gamma/2}{\left[\hbar\omega -
 \left(E_{\mathrm{f}}-E_{\mathrm{i}}\right)\right]^2
  + \left(\Gamma/2\right)^2} ,
    \label{sweight} 
\end{eqnarray} 
corresponding to a phenomenological broadening of the discrete
spectrum  of the polygonal ring.

\begin{figure}
\includegraphics[width=0.48\textwidth,angle=0]{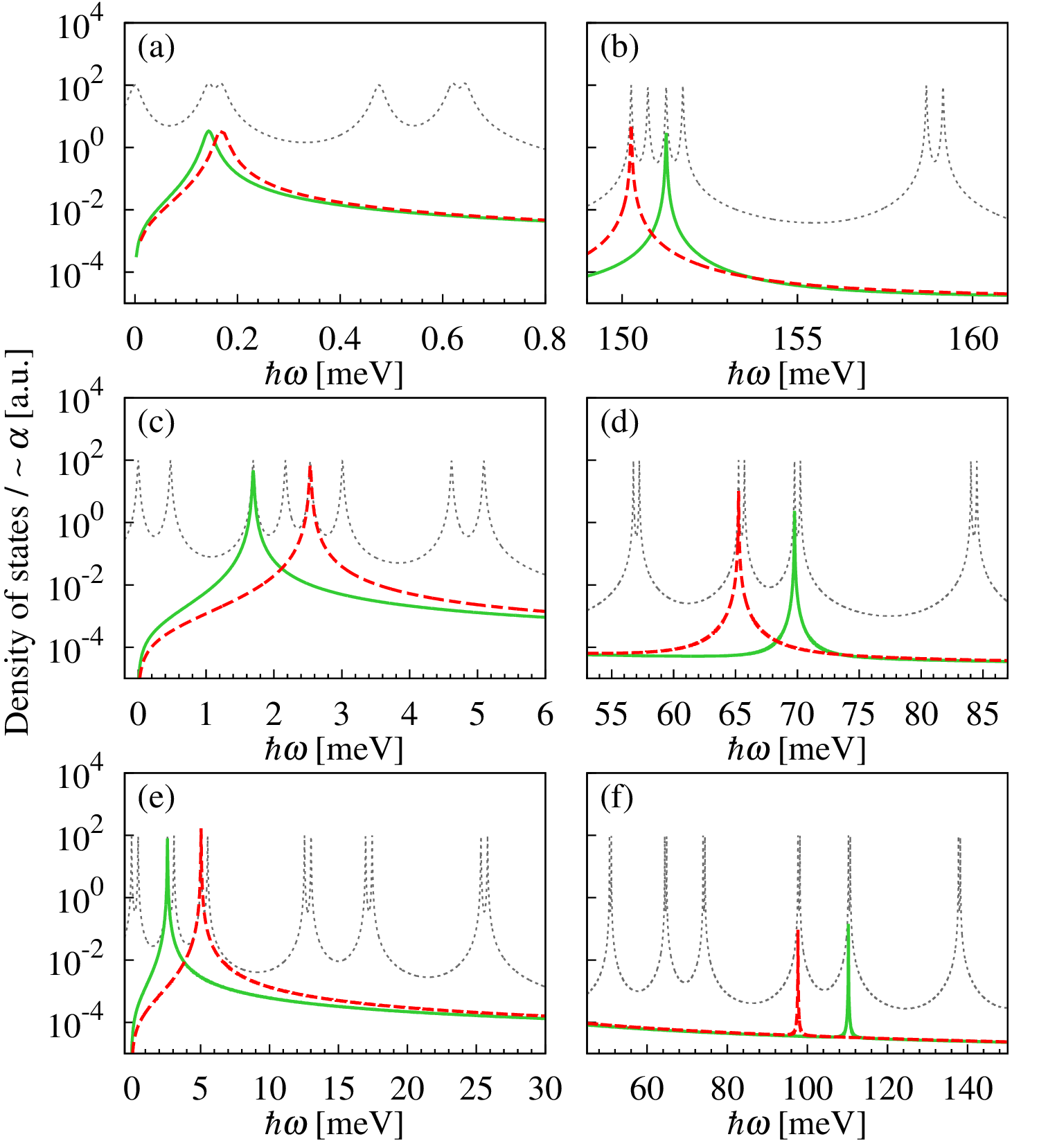}
 \caption{The optically allowed transitions from the ground-state in the 
          presence of clockwise (green, solid) and counterclockwise 
          (red - dashed) polarization superimposed on the density of 
          states from Eq.\ (\ref{sweight}) with $\Gamma=0.056$ meV 
          (gray - dotted), with the ground-state energy tuned to zero for 
          triangular (a) and (b), square (c) and (d), and hexagonal 
          (e) and (f) symmetric rings with sharp corners in a weak 
          perpendicular magnetic field ($0.53$ T). (a), (c), and (e) 
          correspond to transitions to corner-localized states and 
          (b), (d), and (f) transitions to side-localized states. For visibility we 
          use a logarithmic scale for the absorption functions.}
 \label{Fig_abs_1}
\end{figure}


\begin{figure}
\includegraphics[width=0.48\textwidth,angle=0]{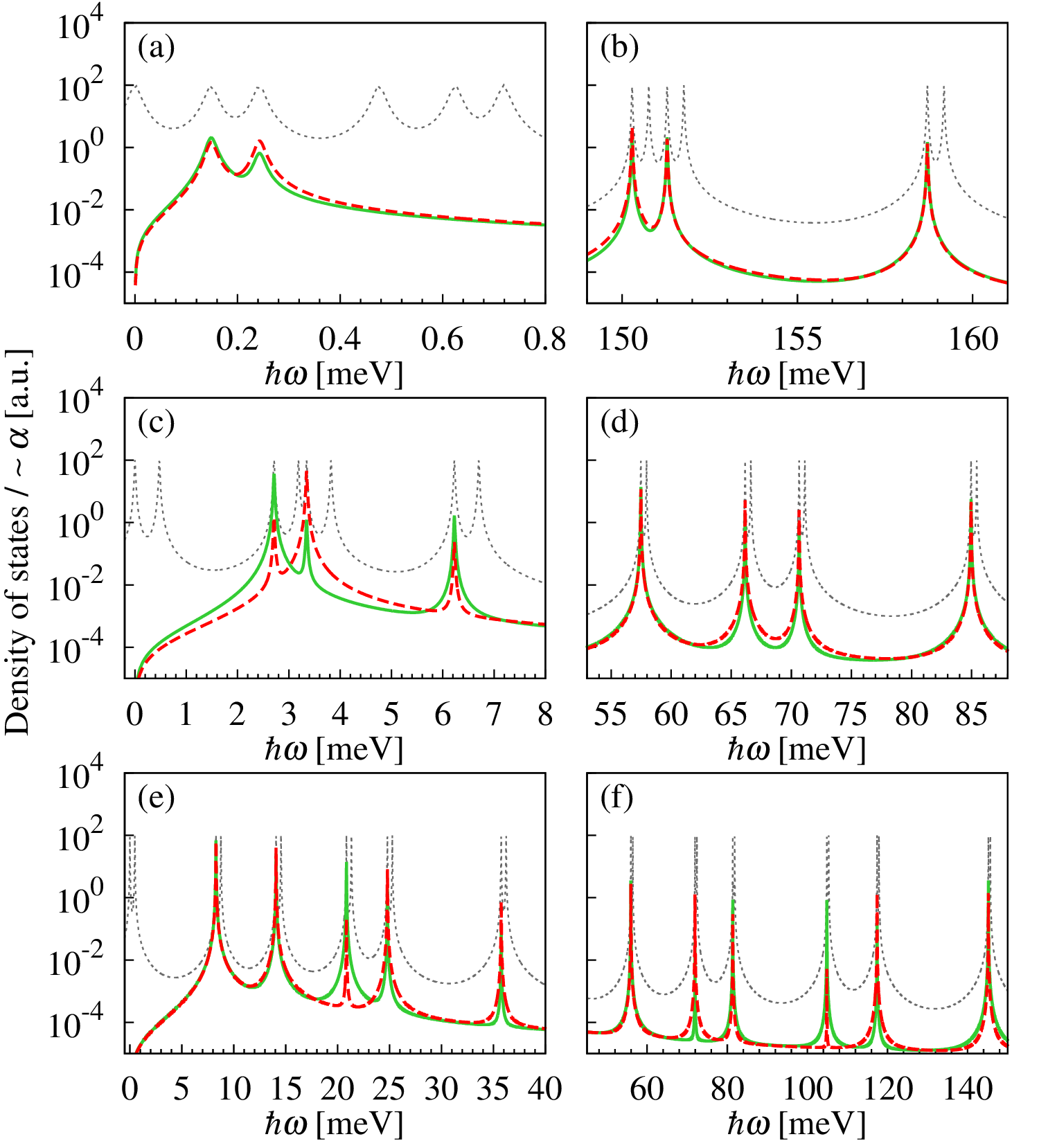}
 \caption{As in Fig.\ \ref{Fig_abs_1} but in the presence of an electric 
          field applied in the ring plane such that it is parallel to 
          one or two ring sides 
          \red{[(a) and (b) $E\approx 0.0056$ mV/nm, $\varphi=\pi/6$,
           (c) and (d) $E\approx 0.11$ mV/nm, $\varphi=0$,
           (e) and (f) $E\approx 0.56$ mV/nm, $\varphi=0$].}
          For visibility we use a logarithmic scale for the absorption functions.}
 \label{Fig_abs_2}
\end{figure}


The localization properties discussed in Sec. \ref{sec:estates} govern 
the optical absorption through the dipole matrix elements 
$\langle f|\bm{\varepsilon}\cdot\bm{d}|i\rangle$, which depend on the
shapes of the wave functions corresponding to states  $|i\rangle$ and $|f\rangle$. 
For simplicity we consider
a weak ($0.53$ T) magnetic field perpendicular to the ring plane which
lifts both degeneracies (due to spin and rotation), but does not 
considerably affect the electron localization. The chosen magnetic field
produces a Zeeman splitting of 0.48 meV.  Since we do not include a 
spin-orbit interaction optical transitions may occur only between states with 
the same spin.  We restrict the investigation to two groups of states, 
corner-localized states and the group consisting of the same number of 
states above them.  For symmetric samples with sharp corners the latter 
states are purely or mostly side-localized 
[Figs.\ \ref{Fig_prob_sym_tr}(c) and \ \ref{Fig_prob_sym_tr}(d) for a 
triangular sample]. We assume that the system is initially in the ground 
state, that is, one electron occupies the lowest energy level and we assume 
a broadening parameter $\Gamma=0.056$ meV.

In Fig.\ \ref{Fig_abs_1} we compare the absorption spectrum of symmetric
triangular [Figs.\ \ref{Fig_abs_1}(a) and\ \ref{Fig_abs_1}(b)], square
[Figs.\ \ref{Fig_abs_1}(c) and\ \ref{Fig_abs_1}(d)], and hexagonal [Figs.\
\ref{Fig_abs_1}(e) and\ \ref{Fig_abs_1}(f)] samples with sharp corners.
We plot all energy intervals between the ground-state and the corner-
[(a), (c), and (e)] or side-localized [(b), (d), and (f)] states,
respectively, on which we superimpose the optical absorption coefficient
calculated according to the formula (\ref{absorption_coeff}) for an
electromagnetic wave circularly polarized in the $x$-$y$ plane. In principle, 
for a triangular ring two transitions to the corner-localized and three 
transitions to the states above the energy gap should be observed. As can 
be seen, both transitions to the lowest-state domain occur, but each one 
is coupled with a different orientation of the photon polarization. The 
reason is that the magnetic field, which points along the positive 
$z$ direction, creates an orbital splitting of the first two excited states. 
The lower of them rotates clockwise in the $x$-$y$ plane, whereas the 
higher rotates counterclockwise. Although three out of the six states shown in 
Fig.\ \ref{Fig_abs_1}(b) have the same spin as the initial state, only two 
optical transitions are observed to states which in the absence of a magnetic 
field would belong to the first, fourfold degenerate energy level above 
the energy gap. As in the previous case each transition is observed in the 
presence of only one polarization direction. The same tendency, i.e., 
coupling of the ground-state (twofold degenerate at zero magnetic field) 
to one of fourfold degenerate states (at zero magnetic field) occurs 
also for transitions to 
higher states (not shown). Also, if an electron is initially in a state from 
the group of fourfold degenerate states then for one polarization orientation 
it may be excited to the twofold degenerate states and for the other
polarization to a fourfold degenerate one. Since for the analyzed triangular sample 
energy separations between corner-localized states are on the order of tens of 
meV and the energy distance from the ground-state to the side-localized states 
ranges from 150 to 160 meV, thus excitation of the ground-state to one of the 
corner-localized states requires absorption of microwave photons, while 
transitions to the side-localized states occur in the presence of a near-infrared 
field. This means that one sample may absorb electromagnetic waves with 
wavelengths differing by orders of magnitude.

Samples with more corners, in principle, could be expected to allow more 
transitions because there are more states with the same spin orientation (four for a 
square and six for a hexagon in each domain). But as shown in 
Figs.\ \ref{Fig_abs_1}(c), \ \ref{Fig_abs_1}(d), 
\ \ref{Fig_abs_1}(e), and \ \ref{Fig_abs_1}(f), still only two transitions 
in each state group occur. The absorption coefficient for transitions to 
corner-localized states increases with the number of corners, while the 
ratio between its value for transitions to side-localized states and 
transitions to corner-localized states rapidly decreases from over $1$ for 
triangular rings to values on the order of $10^{-3}$ for hexagonal samples.
Moreover, the magnitude of the absorption coefficient depends on the 
polarization type.  In the absence of an external electric field it is 
usually higher for counterclockwise-polarized light. The splitting of the 
dipole-active absorption peak into mainly two peaks with a growing magnetic field 
is a well-known phenomena for quantum dots of various shapes 
\cite{Gudmundsson91, Krahne01, Magnusdottir99}. The same can be stated about 
the opposite trends for the evolution of the height of the two absorption 
peaks with increasing magnetic field.

An external electric field may change the picture, as is shown in 
Fig.\ \ref{Fig_abs_2} where the field is applied in the ring plane. For both
polarization directions all spin-allowed transitions take place, but with 
different values of the absorption coefficient, which shows that optical
experiments may be used to probe the sample geometry. The electric field
strength which allows all transitions to be "opened" increases with the number
of corners.  Triangular samples require weak fields because their corners
are relatively well separated, while corners of a hexagon defined by the
same radius are much closer to each other and thus an electric field of
the same value only slightly changes the height of the localization peaks. 
The dipole matrix elements depend on the symmetry of the wave function and may 
vanish for some pairs of states, as in the case of transition to the fifth state in 
Fig.\ \ref{Fig_abs_1}(b).  But the electric field breaks the wave function
symmetry (and shifts the energy levels), and thus it also changes the matrix
elements and opens some other transitions, like the one from the ground
to the fifth state [Fig.\ \ref{Fig_abs_2}(b)]. When the electric field
is strong enough to induce localization in single corner areas, then 
transitions to side-localized states are much more pronounced than those 
to corner-localized ones.
The same situation, for the same symmetry reasons, is observed for the non
symmetric sample shown in Fig.\ \ref{Fig_prob_non_tr}. The absorption
spectrum is also sensitive to the angle which the field forms with the
$x$ axis; rotation of the field may open, close, or change the strength 
of some transitions.


\section{\label{sec:conclusions} Conclusions}

We studied spectral and optical properties of 2D polygonal quantum rings. 
We showed that the polygonal geometry induces two- and fourfold degeneracies
and formation of an energy gap which depends on the number of corners and 
the lateral side thickness.  In general the lowest energy states are localized 
in corner areas, forming a low-energy shell. The probability density is very 
sensitive to the ring shape.  Even if the geometry of the sample only slightly 
differs from a regular ring, the electron localization becomes strongly 
nonuniform around the polygon. The charge carriers in the ground-state are 
always localized in the corner with the largest area. A certain softening of 
the corners changes the electron localization only of higher energy-states.  
The localization pattern may be, to some extent, controlled by an external 
electric field, which may change the effective potential wells associated with 
corners. This may also result in breaking the symmetry of regular polygons 
as inducing a symmetric probability distribution in nonsymmetric samples.

In order to predict basic optical properties related to the corner localization,
we calculated the absorption coefficient using the linear response method. We 
did not include spin-orbit interaction, and thus optical transitions occur only 
between states with the same spin. Other selection rules are related to the 
symmetry of the wave functions. Some transitions are forbidden and others are 
allowed depending on the (circular right or left) photon polarization. In the 
absence of an external electric field only two transitions from the ground-state 
to the next higher corner-localized states and side-localized states occur. We 
showed that, 
\red{as in to long triangular core-multishell wires \cite{Gradecak05,Qian04,Qian05,Qian08}, triangular rings interact 
with radiation from different domains, possibly microwave and near-infrared at the same time.}
Since the external electric field
changes the wave function geometry, it also affects the absorption coefficient through 
the dipole matrix elements and ``opens'' previously ``closed'' 
transitions, blocks others or 
changes their intensity
\red{, i.e., allows contactless control of optical properties.}


\begin{acknowledgments}
This work was financially supported by the Research Fund
of the University of Iceland, the Nordic network NANOCON-
TROL, project No.: P-13053, and by MINECO-Spain (Grant
No. FIS2011-23526).

\end{acknowledgments}


\bibliographystyle{apsrev4-1}

\end{document}